\documentstyle[12pt,a4,epsf]{article}
\begin{document}
\baselineskip=18pt


\newcommand{\greeksym}[1]{{\usefont{U}{psy}{m}{n}#1}}
\newcommand{\umu}{\mbox{\greeksym{m}}}
\newcommand{\udelta}{\mbox{\greeksym{d}}}
\newcommand{\uDelta}{\mbox{\greeksym{D}}}
\newcommand{\uPi}{\mbox{\greeksym{P}}}
\newcommand{\hdmo}{HEIDELBERG-MOSCOW experiment~}
\newcommand{\bb}{$\beta\beta$}
\newcommand{\nbb}{$2\nu\beta\beta$}
\newcommand{\ncbb}{$2\nu\chi\beta\beta$}
\newcommand{\ch}[2]{$\rm ^{#1}#2 $}
\newcommand{\lan}{\left\langle}
\newcommand{\ran}{\right\rangle}
\newcommand{\lmi}{\left|}
\newcommand{\rmi}{\right|}
\newcommand{\lro}{\left(}
\newcommand{\rro}{\right)}
\newcommand{\lre}{\left[}
\newcommand{\rre}{\right]}
\newcommand{\nn}{\nonumber}

\def \beq {\begin{equation}}
\newcommand{\etal}{\textit{et. al.}}

\def \gs  {Gran Sasso Underground Laboratory} 
\def \onbb {$0\nu\beta\beta$~ }
\def \tnbb {$2\nu\beta\beta$~ }
\def \mnbb {$0\nu\chi\beta\beta$~ }
\def \bb {$\beta\beta$~ }


%
\title{\bf Critical View to "The IGEX neutrinoless double beta decay
experiment...."
            published in Phys. Rev. D, Volume 65 (2002) 092007}

\author{H.V. Klapdor-Kleingrothaus$^{1,3}$, A. Dietz$^{1}$, 
I.V. Krivosheina$^{1,2}$\\
%
\it $^{1}$Max-Planck-Institut f\"ur Kernphysik,\\
        Postfach 10 39 80, D-69029 Heidelberg, Germany\\
\it     $^{2}$Radiophysical-Research Institute, Nishnii-Novgorod, Russia\\
\it     $^{3}$Spokesman of the HEIDELBERG-MOSCOW,\\ 
\it     HDMS and GENIUS(TF) Collaborations,\\ 
        e-mail: klapdor@gustav.mpi-hd.mpg.de,\\
         home page: http://www.mpi-hd.mpg.de/non$\_$acc/}

\maketitle              
\vskip.5cm

{\bf Abstract.} 
        Recently a paper entitled ``The Igex $^{76}${Ge} 
        Neutrinoless Double-Beta Decay Experiment: 
        Prospects for Next Generation Experiments'' 
        has been published 
        in {\it Phys. Rev. D 65 (2002) 092007}
\cite{Noch-Dum-Av02}.
        In view of the recently reported evidence 
        for neutrinoless double beta decay 
\cite{KK-Evid01,KK02-Found,KK02-PN}
        it is particularly unfortunate that the IGEX paper 
        is rather incomplete in its presentation. 
        We would like to point out in this Comment that and why 
        it would be highly desirable to make more details about 
        the experimental conditions and the analysis of IGEX 
        available. 
        We list some of the main points, 
        which require further explanation. 

        We also point to an arithmetic mistake in the analysis 
        of the IGEX data, the consequence of which are too high half life 
        limits given in that paper.


\vskip0.4cm
\noindent
{\bf Pacs: 23.40.Bw} (Weak interaction and lepton (including neutrino) 
aspects).\\
{\bf Pacs: 23.40.Hc} (Relation with nuclear matrix elements and nuclear 
structure)\\
\vskip3mm
\noindent
{\bf keywords}: weak interaction, double-beta decay, matrix elements, neutrinos

\vskip0.6cm
        Recently a paper entitled 
        "The Igex $^{76}${Ge} Neutrinoless Double-Beta Decay 
        Experiment: Prospects for Next Generation Experiments'' 
        has been  published in 
        {\it Phys. Rev. D 65 (2002) 092007}. 
        In view of the recently reported evidence 
        for neutrinoless double beta decay 
\cite{KK-Evid01,KK02-Found,KK02-PN,Repl-KK,Backgr-NIM-03,KK-StBr02}
        it is unfortunate that the IGEX paper is rather incomplete 
        in its presentation. 
        It would be highly desirable if more details about 
        the experimental conditions and the analysis of IGEX 
        would be made available. 
        In this Comment we list some of the main points, 
        which require further explanation. 
        We also point to an arithmetic mistake in the analysis 
        of the IGEX data, the consequence of which are too high half life 
        limits given in that paper.
\\

        Before we go into details, some general points of IGEX should be
        clarified.
        The IGEX double beta experiment stopped operation already in 1999
\cite{Kirp00}.
        Consequently the authors in 
\cite{Noch-Dum-Av02}
        show the analysis, which they
        showed already at the NANP 99 Conference in Dubna and published in the
        Proceedings of that conference
 \cite{DUM-RES-AVIGN-2000}.\\

        Some general, and fundamental information about the IGEX 
        experiment is missing. 
        The paper does not give sufficient detail on the history, 
        quality, stability and run times of the detectors.
        Also, for example the small 'duty cycle' of the experiment 
        is not explained. 
        The background reached in the experiment is even not mentioned. 
        The statistical methods of analysis are not described. 
        No analysis of the background lines has been published, 
        and no Monte Carlo simulation of the background is presented.
        No spectrum is shown over the full energy range. 
\\

        IGEX working with in total 9\,kg of enriched 76 Germanium, 
        collected in 8-9\,years of operation in total {\it only} 
        {\bf 117\,molyears} of data. 
        This corresponds to {\bf 8.7\,kg\,years} - 
        which propably means that the 
        IGEX experiment took data only in a short part 
        of its time of operation (- or only a small part of the data 
        was selected for analysis). 

        We shall comment the following topics:

\section{Measured Spectrum}

        The authors do not show in 
\cite{Noch-Dum-Av02}
        the measured spectrum over the full energy range,
        so they give no feeling for experimental parameters like energy
        resolution, stability of the electronics, and understanding of their
        background. Only in an earlier publication 
\cite{Aals99}
        they show a full spectrum, but compressed 
        to {\bf 10\,keV per channel}. 
        This is by far not adequate to measurements of spectra 
        with Germanium semiconductor detectors. 
        Competitive experiments present their data in 0.36\,keV or 
        at least 1\,keV per channel.  
        They further do not show any 
        identification of background lines (except 2 lines).

\section{Pulse Shape Analysis}

The method of pulse shape analysis (PSA) used in that 
        paper seems not yet to be a technically mature procedure. 
        It makes among others use of a visual determination 
        of the shape of the pulses
\cite{DUM-RES-AVIGN-2000,Noch-Dum-Av02}.
        This casts doubt on the reliability of the background determination.
        It is questionable how in such a way quantitative results can be
        obtained. It seems unavoidable that in such a way the authors
        naturally run into the danger of producing too sharp limits for the
        half life.
        It is called a 'rudimentary PSD technology' by
        the authors themselves 
\cite{Aals99}.
        The data analyzed with PSA contain about 52\,molyears of data,
        corresponding to 3.9\,kg\,years.\\ 


\section{Statistical Analysis and Background}

        The experiment collected 117\,molyears (8.7\,kg\,years) 
        of data, of them 3.9\,kg\,y with 'visual' PSA.
        {\it The background in the IGEX experiment is not mentioned in 
\cite{Noch-Dum-Av02}, 
        and also not in the NANP Proceedings 
\cite{DUM-RES-AVIGN-2000}}. 
        It is said in 
\cite{Aals99}
        to be ~0.2\,counts/kg\,y\,keV 
        for  p a r t of the data (i.e. usually higher).
        The authors do not say in the paper 
\cite{Noch-Dum-Av02}
        (and also not in the NANP Proceedings 
\cite{DUM-RES-AVIGN-2000}), 
        how they analyze their data in the range of the
        potential neutrinoless double beta decay signal. 
        This is a crucial point. 
        They talk only about standard statistical techniques. 
        Since there are many standard techniques, 
        this makes it difficult to judge the significance of their result.

        Furthermore, the usual procedure recommended 
        by the Particle Data Group 
\cite{RPD00,PD96}
        in the case that the countrate is smaller than the
        expected background rate (which is the case in their spectrum, see
        their Fig. 2), is, to give also the more conservative value obtained
        when setting the count rate equal to the expected background rate. 
        This would correspond to the 'sensitivity' 
        of the experiment according to Feldman-Cousins 
\cite{Feldm01}. 

        This has  n o t  been done by the authors of 
\cite{Noch-Dum-Av02}.  

        In the present paper 
\cite{Noch-Dum-Av02}
        it is announced without giving any details, 
        that 'standard statistical techniques' lead - 
        for the PSA (Pulse Shape Analysis) data - to a
        limit of $1.57\times{10}^{25}$ years. 
        {\it This value is obtained by mistake.} 
        What the authors do, is that in their eq. (5),  
        which in general reads (see 
\cite{Aals99})

        \centerline {$T^{0\nu}_{1/2} > {(N \cdot t \cdot \ln{2})/{C},}$}

\noindent 
        they insert by mistake (the same mistake is found in their paper 
\cite{DUM-RES-AVIGN-2000})
        for $N \cdot t$ the number of mol years of the {\it full experiment} 
        (117 mol\,years $\cdot \ln{2} \equiv 4.87 \times {10}^{25}$\, y) 
        but in the denominator they choose the value 
        for the 90\% confidence limit of the number of events 
        attributable to \onbb decay, $C$, equal to the one  
        which is valid for the {\it Pulse-Shape analyzed} spectrum, 
        namely $C = 3.1$. 
        Instead, the latter value should be around $C = 4.5$ 
        for the full spectrum. 

        The half-life deduced from the full data then would be 
        $T^{0\nu}_{1/2} < 1.1 \cdot {10}^{25}$\, y, 
        as stated correctly in their earlier paper 
\cite{Gonz00} 
        (where they give 
        $T^{0\nu}_{{1/2}} < 1.13 \cdot {10}^{25}\,y$)  
        instead of the given $T^{0\nu}_{1/2} < 1.57 \cdot {10}^{25}$\, y. 
        The Feldman-Cousins (FC) 
\cite{Feldm01}
        sensitivity of the experiment is 
\begin{eqnarray}
        T^{0\nu}_{{1/2}~{(Full)}} < 0.52 \cdot {10}^{25}\,y.
\end{eqnarray}
        For the PSA spectrum the value to be inserted into 
        the numerator should 
        be 

\noindent
        52.51\,mol\,years$\cdot \ln{2} = 2.2 \cdot {10}^{25}$y, 
        which yields with their $C = 3.1$ for the half-life limit 
        
        \centerline{$T^{0\nu}_{{1/2}~{(PSA)}} < 0.71 \cdot {10}^{25}$y.} 

        Here it should be noted that $C = 3.1$ is depending 
        on the width of the energy range analyzed. 
        If this energy range is increased by only 20\%, $C$ 
        will become 3.8 and 

        \centerline{$T^{0\nu}_{{1/2}~{(PSA)}} = 0.58 \times 10^{25}$y.}

        The FC sensitivity in this case is 
\begin{eqnarray}
        T^{0\nu}_{{1/2}~{(PSA)}} < 0.28 \cdot {10}^{25}\,y.
\end{eqnarray}
        These corrected estimates of the half life limits from the 
        IGEX data 
        correspond more naturally  to those deduced from an experiment
        having almost one order of magnitude higher statistics 
\cite{HDM01}, 
        which yielded a limit of $1.3\times{10}^{25}$\,years, 
        from the full data of 
        a statistical significance of 53.9\,kg\,y.


\section{The Effective $\nu$ Mass}

        Starting from their incorrectly determined 
        half-life limit the authors claim 
        a range of the effective neutrino mass of (0.33 - 1.35)\,eV. 

        These numbers given by 
\cite{Noch-Dum-Av02}
        and already earlier in 
\cite{DUM-RES-AVIGN-2000} 
        unfortunately have been uncritically cited 
        in several theoretical papers (see e.g.
\cite{Citat02}).

        The effective neutrino mass limit $\langle m \rangle$ 
        deduced from the half life limit
        should read for different matrix elements correctly, 
        as given in Table 1.
        It is seen that the numbers deducible from IGEX are almost a
        factor of 2 larger than reported by the authors of 
\cite{Noch-Dum-Av02}.

\begin{table}[h]
\begin{center}
\newcommand{\m}{\hphantom{$-$}}
\renewcommand{\arraystretch}{.5}
\setlength\tabcolsep{1.7pt}
\begin{tabular}{c|c|c|c|c}
\hline
        &       $\langle m \rangle$\,eV &       $\langle m \rangle$\,eV from:
&       $\langle m \rangle$\,eV from:   &       $\langle m \rangle$\,eV from:\\

        & Positive      &       {\footnotesize Latest results}  
&       our     &       Aalseth et al.\\        
        Model   &       Evidence        
& {\footnotesize (2001)\cite{HDM01}}    &       (conservative)  
&       {\footnotesize PRD 65 (2002)}   \\
                &       \cite{KK-Evid01,KK02-Found,KK02-PN}     
&       {\footnotesize HEIDELBERG-}&    analysis of     
&       {\footnotesize 092007}\\
        &       (best value)    & {\footnotesize -MOSCOW 90\%c.l.}      
&       {\footnotesize Aalseth et al.}  & 90\% c.l.\\
                &       ${\rm T}_{1/2}^{0\nu} = $       
&       (full data)
&       {\footnotesize  data (90\% c.l.)}       &       claimed\\
                &       $1.5 \times 10^{25}~ y$ &       $1.3\times{10}^{25}$\,years
&       $0.5\times{10}^{25}$\,years&\\
\hline
\hline
&&&\\
Weak Coupling   &       0.34    & $<$ 0.37      & $<$ 0.59      & $<$ 0.33\\
Shell Model     \cite{Hax84}    &                       &       
&       \\
\hline
&&&\\
        QRPA \cite{Eng88}       &  1.37 &       $<$ 1.47        &       $<$ 2.40        &       $<$ 1.35\\
\hline
&&&\\
        QRPA \cite{Vog-Trento}  &  0.97 &       $<$ 1.04        &       $<$ 1.67        &       $<$ 0.94\\
\hline
&&&\\
        QRPA \cite{Civit-Tom87} &  0.39&        $<$ 0.42        &       $<$ 0.69        &       $<$ 0.39\\      
\hline
&&&\\
        QRPA \cite{Klapdor89-90}        &  0.39 &       $<$ 0.42        &       $<$ 0.69        &       $<$ 0.39        \\
\hline
&&&\\
Large Scale &&&\\
Shell Model     \cite{Caur96}   &  1.07 &       $<$ 1.15        &       $<$ 1.86        &       $<$ 1.05\\      
\hline
&&&\\
RQRPA \cite{Faes01}     &  0.53 &       $<$ 0.57        &       $<$ 0.92        &       $<$ 0.52\\
\hline
&&&\\
SQRPA  \cite{St-KK01-1,St-KK01-2}       & 0.44--0.52    
&       $<$(0.47--0.56) &       $<$(0.78--0.90)
&       $<$(0.44--0.51)\\
\hline
\end{tabular}
\end{center}
\caption{\label{Matr-Elem}The effective neutrino mass limits  deduced 
        from the half-life limits with different matrix elements.
        Also shown are the neutrino masses deduced from the best 
        value of ${\rm T}_{1/2}^{0\nu} = 1.5 \times 10^{25} 
        {\rm~ y}$ determined in 
\protect\cite{KK-Evid01,KK02-Found,KK02-PN}. 
}
\end{table}


        For comparison we give in Table 1 the values corresponding to the
        half life limit of $1.3\times{10}^{25}$\,years  
        (90\% c.l.) by the \hdmo with the full data taken in 53.9\,kg\,y 
\cite{HDM01}.
        They are consistent with the present claim 
\cite{KK-Evid01,KK02-Found,KK02-PN}
        of an effective mass with best value of 0.39\,eV
\cite{KK02-Found}.

        \section{Nuclear Structure}

        The discussion of nuclear structure and matrix 
        elements is incomplete and seems superficial. 
        It ignores recent work (after 1996).
        We just refer to 
\cite{Faes01},\cite{St-KK01-1},\cite{St-KK01-2}.
        It carries along calculations which are known to have deficiencies,
        e.g. the QRPA calculations of ref. 
\cite{Eng88,Vog-Trento} 
        which they mix in their Ref. [25]  
        with a paper not yielding any information 
        about \onbb matrix elements at all. 
        These calculations suffer 
        from not using a realistic nucleon-nucleon potential.  
        It also carries along the weak-coupling limit shell model (ref. 
\cite{Hax84}) which contains a
        by far too low configuration space. They carry along further
        the socalled large-scale shell model  
\cite{Caur96}, 
        which as result of
        its limited configuration space does not fulfill the Ikeda sum rule
        (see, e.g. the review
\cite{Faes98}) 
        and consequently systematically
        underestimates the matrix element (leading
        to a corresponding overestimate of  the effective neutrino mass).
        It might be mentioned, that the calculation of 
\cite{Klapdor89-90} 
        gave the {\it prediction} most close to the experimental \tnbb 
        decay half-life of $^{76}{Ge}$ 
        of $(1.74 ^{+0.18}_{-0.16})\times{10}^{21}$\,years  
\cite{HDM97,KK-doerr03}.
        It underestimates the 2$\nu$ matrix element by only 31\%.

        \section{Conclusion}

        Summarizing, 
        it is unfortunate - particularly in view 
        of the recently reported evidence 
        for neutrinoless double beta decay 
\cite{KK-Evid01,KK02-Found,KK02-PN,Repl-KK,Backgr-NIM-03,KK-StBr02} - 
        that the IGEX paper - apart from the 
        too high half-life limits presented, 
        as a consequence of an arithmetic error - 
        is rather incomplete in its presentation. 
        It does not give sufficient information on 
        the experimental conditions 
        and the analysis to judge the significance 
        of their given results. 
        It would be highly desirable if some basic points discussed 
        in this Comment would be clarified and made available 
        to scientific discussion.


\end{document}